\documentclass[a4paper,twocolumn,11pt,accepted=2022-09-23]{quantumarticle}
\pdfoutput=1
\usepackage[utf8]{inputenc}
\usepackage[english]{babel}
\usepackage[T1]{fontenc}
\usepackage{amsmath}
\usepackage{hyperref}
\usepackage[numbers, sort&compress]{natbib}

\newcommand{\ket}[1]{\vert {#1}\rangle} 
\newcommand{\abs}[1]{\vert#1\vert}

\begin{document}
\title{Exploring entanglement resource in Si quantum dot systems with operational quasiprobability approach}

\author{Junghee Ryu}
\affiliation{Division of National Supercomputing, Korea Institute of Science and Technology Information, \\Daejeon 34141, Republic of Korea}
\orcid{0000-0002-1941-9923}
\author{Hoon Ryu}
\email{elec1020@kisti.re.kr}
\affiliation{Division of National Supercomputing, Korea Institute of Science and Technology Information, \\Daejeon 34141, Republic of Korea}
\orcid{0000-0001-9302-2759}
\maketitle

\begin{abstract}
We characterize the quantum entanglement of the realistic two-qubit signals that are sensitive to charge noises. Our working example is the time response generated from a silicon double quantum dot
(DQD) platform, where a single-qubit rotation and a two-qubit controlled-NOT operation are conducted sequentially in time to generate arbitrary entangled states. In order to characterize the entanglement
of two-qubit states, we employ the marginal operational quasiprobability (OQ) approach that allows negative values of the probability function if a given state is entangled. While the charge noise, which is
omnipresent in semiconductor devices, severely affects logic operations implemented in the DQD platform, causing huge degradation in fidelity of unitary operations as well as resulting two-qubit states,
the pattern in the OQ-driven entanglement strength turns out to be quite invariant, indicating that the resource of quantum entanglement is not significantly broken though the physical system is exposed
to noise-driven fluctuations in exchange interaction between quantum dots.
\end{abstract}

\section{Introduction}
\label{sec:intro}
Quantum correlations have been regarded as key ingredients for advantages of several applications in quantum information technologies such as quantum computing, quantum communications, quantum metrology,
and so on. These resources enable the quantum advantages in ways that classical methods can do only with much less efficiency or cannot. Quantum entanglement, the most celebrated example of the quantum
resources, plays a crucial role in quantum information applications \cite{Horodecki09,ReviewBell}. For example, in quantum communication, a shared entanglement between two parties can serve as a channel for
teleportation of quantum bits (qubits) \cite{Teleportation}. Quantum algorithms being represented by the most well-known one proposed by Shor \cite{ShorAlgorithm} that enables factorization of large integers much
faster than the best classical algorithm known so far, are not feasible without quantum entanglement. It has been also shown that quantum entanglement can improve a parameter estimation in sensing applications
\cite{Changhyoup21}. Recently, significant advances for physical realization of universal quantum computers have been demonstrated using various platforms such as superconductors, trapped ions, silicon (Si)
quantum dots (QDs), and so on \cite{Arute19,Mooney19,Pogorelov21,Debnath16,Wright19,Watson18}. Especially, Si QDs have great potential for designs of scalable quantum circuits since they can be fabricated
with industrial-standard lithographical processes and can provide extremely long spin-coherence time \cite{Steger12,Tyryshkin12,Veldhorst14,Veldhorst15}. The recent work done by Zajac {\it et al.} has demonstrated
the successful realization of a fast Controlled-NOT (CNOT) gate with a single-step control pulse in a Si double QD (DQD) platform \cite{Zajac18}, where the CNOT logic is a two-qubit entangling block that is essential
for designs of universal quantum computing.

Fidelity is the physical quantity that has been widely employed to determine the preciseness of quantum logic operations and corresponding output states. But, the fidelity itself cannot determine whether the output
state is entangled or not; instead it quantifies the similarity of the obtained outcome and the desired one. To quantify the entanglement strength of an unknown state, one may employ the entanglement measures
\cite{Horodecki09,Otfried09}. The entanglement measures, however, require all the elements of a state density matrix that can be obtained only with a full quantum state tomography, and therefore are computationally
expensive in general. For the simplest case of a single qubit, the density matrix is expressed by a 2$\times$2 complex matrix, and it can be characterized by four parameters known as the Stokes parameters. For
an $n$-qubit state, the size of parameters that must be determined is sharply increased to 4$^n$. There have been many studies for analytical derivation of the entanglement measures, but so far only the case of
two-qubit systems has been well established. In addition, some measures are not linked to directly measurable quantities as they need to involve additional mathematical operations to verify the entanglement, so
they may not be appropriate for physical implementation. For examples, the measures such as $negativity$ and $concurrence$ require a diagonalization or a partial transpose operation that
are not easy to be implemented in laboratories.

In the field of quantum optics, various nonclassical effects have been analyzed by the quasiprobabilities such as Wigner, Glauber-Sudarshan, and Q functions~\cite{Wigner32, Husimi40, Glauber63, Sudarshan63, Cahill69}. Due to the uncertainty principle, these functions can have negative values which are regarded as indicators of the nonclassical phenomena. However, an operational interpretation of such negative values is unclear, wherein preparation, operation, and measurement processes cooperate explicitly~\cite{Ferrie2011}. There have been many studies to investigate operationally defined negative probabilities to characterize the nonclassical features including the quantum entanglement~\cite{Jiyong2015, Sperling2018, Sperling2020, Bohmann2020, Jiyong2021}.

Recently, an alternative method of entanglement verification, called as the $operational$ $quasiprobability$ (OQ), has been proposed. Employing a probability function that can generate negative values,
the OQ method characterizes various quantum natures of a given logic state with negative probabilities that are fundamentally impossible to be obtained with classical probability density functions \cite{Ryu13,Jae17}.
In principle, the OQ method is advantageous over the entanglement measures in a sense that the probability function can be determined only directly measurable quantities that can be easily tested in laboratory \cite{Ryu19}.
Moreover, in general, the OQ method requires less number of measurements for entanglement verification, so the computational cost can be saved compared to the case of entanglement measures that involve
a full-state tomography processes \cite{Otfried09}. Though it has been theoretically understood that the OQ probability function can be solely constructed with directly measurable local observables, however, so far its practicality
for the characterization of entanglement has not been elaborately examined against real physical problems, raising natural interest and motivation for related studies with a focus on entangling logic operations
under realistic conditions. In this work, we explore the utility of the OQ method for characterization of quantum entanglement of two-qubit states that are generated in a Si DQD system which encodes qubits to spins
of confined electrons. The experimental conditions of DQD systems are realistically mimicked with our in-house simulation code package, which has been validated by modeling quantum logic operations implemented
in a physically realized Si DQD platform \cite{Zajac18,Kang21,Ryu22}. With a focus on a simple circuit that sequentially conducts a one-qubit rotation and a CNOT operation, we apply the marginal OQ method to
the time response of a DQD system to quantify the amount of entanglement, and compare the result to the one obtained with an ideal toy model. Entanglement characterization of the DQD system under noisy
conditions is discussed in detail, delivering a non-negligible message that the entanglement resource can be quite robust to charge noises even though the noise-driven degradation in fidelity is clearly observed
for gating operations and output states.

\section{Methods}
\label{sec:pre}
\begin{figure}
	\centering
	\includegraphics[width=\columnwidth]{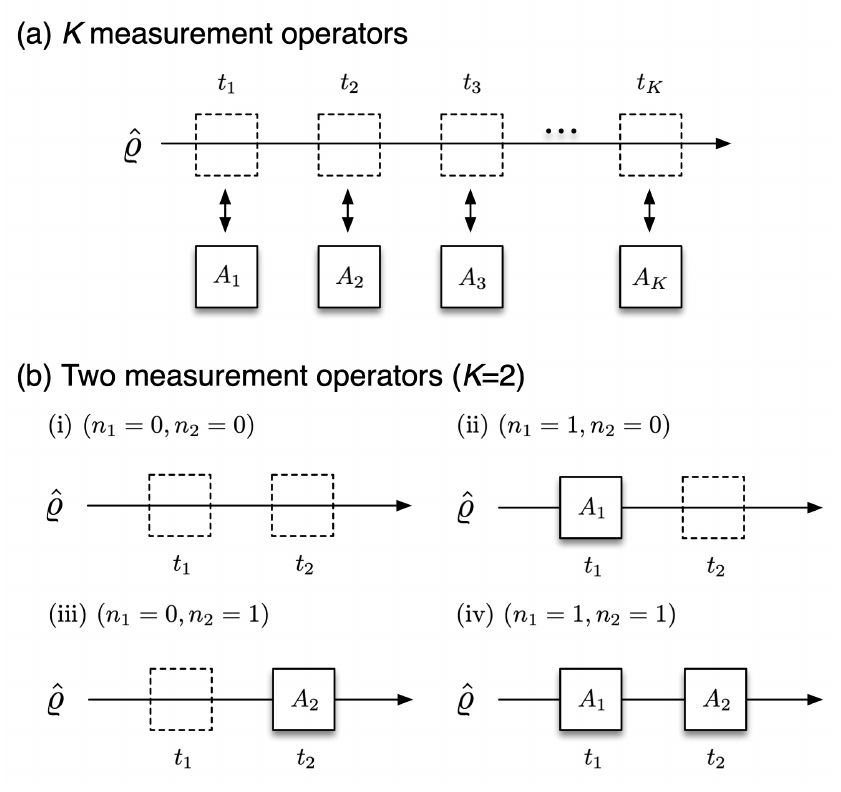}
\caption{
	\textbf{Schematic diagram of measurement configurations.} (a) When $K$ possible measurement operators $A_k$ are selectively and sequentially considered at time $t_k$, $2^K$ measurement configurations
	can be constructed. (b) For two operators $A_1$ and $A_2$ case, there are four measurement configurations: (i) no measurement, (ii) $A_1$ measurement at time $t_1$, (iii) $A_2$ measurement at $t_2$,
	and (iv) a sequential measurement of $A_1$ and $A_2$ at $t_1$ and $t_2$.
	}
\label{fig:schm}
\end{figure}

Since the cornerstone of our method is the OQ, here we briefly review its mathematical details for handling general multi-qubit systems. Consider that $K$ nondegenerate measurement operators $A_k$ are
selectively and consecutively applied on a qubit, each of which has two possible outcomes denoted by $a_{k} \in \{0,1\}$. Each $A_k$ is measured at time $t_k$ with $t_k < t_l$ for $k < l$, and we consider a
projective measurement described by projectors $\Pi_{k}^{a_k}$ as $A_k = \sum_{a_k} (-1)^{a_k} \Pi_{k}^{a_k}$ (for the case of general measurements, see Ref. \cite{Ryu13}). Depending on the selection of
the operators, one can perform $2^K$ different measurement configurations including void measurement which refers to no selection of any measurements and sequential measurements (see Figure \ref{fig:schm}(b)
for $K=2$). We denote each measurement configuration by $n$-tuple as ${\bf n}=(n_1, \dots, n_K)$ with $n_k \in \{0,1\}$, where $n_k = 1$ if the measurement operator $A_k$ is selected to be measured at
time $t_k$ and $n_k =0$ otherwise.

For each setup, we construct the following form of expectation value:
\begin{eqnarray}
C({\bf n}) = \sum_{{\bf a}} (-1)^{{\bf n \cdot a}} P({\bf a}|{\bf A}_{\bf n}),
\label{eq:characteristic_ftn}
\end{eqnarray}
where ${\bf{a}} = (a_1, \dots,a_K)$ with $a_k \in \{0,1\}$, $\sum_{\bf a} = \sum_{a_1} \cdots \sum_{a_K}$, and ${\bf n \cdot a} = \sum_{k=1}^{K} n_k a_k$. Here $P({\bf a}|{\bf A}_{\bf n})$ is a sequential probability
of the outcome ${\bf a}$ given ${\bf A}_{\bf n}$ measurement setup, where ${\bf A}_{\bf n}$ involves the only measurements corresponding to a nonzero $n_k$. For example, when two measurement operators
$A_1$ and $A_2$ are selected to be measured among $K$ measurements (only $n_1$ and $n_2$ are nonzero of ${\bf n}$), there are four measurement configurations whose expectation values are given by
(i) $C(0,0) = 1$, (ii) $C(1,0) = \sum_{a_1} (-1)^{a_1} P(a_1 | A_1)$, (iii) $C(0,1) = \sum_{a_2} (-1)^{a_2} P(a_2 | A_2)$, (iv) $C(1,1) =\sum_{a_1,a_2} (-1)^{a_1 + a_2} P(a_1, a_2 | A_1, A_2)$ as shown in Figure
\ref{fig:schm}(b). In particular, for the projective measurements the sequential probability can be put as
\begin{eqnarray}
P (a_i, a_j | A_i, A_j)=\text{Tr}[ \Pi_{j}^{a_j} \Pi_{i}^{a_i} \varrho\,(\Pi_{j}^{a_j} \Pi_{i}^{a_i})^{\dagger}].
\end{eqnarray}
Indeed, the formula (\ref{eq:characteristic_ftn}) describes the expectation values for all measurement configurations for the $2^K$ cases.

The OQ function is defined by applying a discrete Fourier transform on $C({\bf n})$ as
\begin{eqnarray}
\mathcal{W} ({\bf a}) \equiv \frac{1}{2^K} \sum_{{\bf n}} (-1)^{-{\bf a} \cdot {\bf n}} C({\bf n}),
\label{eq:quasiprob}
\end{eqnarray}
where $\sum_{\bf n} = \sum_{n_1} \cdots \sum_{n_K}$. The expectation $C(\bf{n})$ can be reproduced by the inverse Fourier transform of $\mathcal{W}({\bf a})$. That is, the OQ function has a full information of the
quantum expectation without loss of the generality. It is a real-valued function and normalized to one over all ${\bf a}$; $\sum_{\bf a} \mathcal{W}({\bf a}) =1$. The sum of $\mathcal{W}({\bf a})$ over a part of ${\bf a}$
reproduces the marginal quasiprobability of the rest and for a single argument $a_k$ it is equivalent to the probability of measuring $a_k$, $\mathcal{W}(a_k) = P(a_k | A_k)$. To quantity the degree of the nonclassicality,
the sum of the negative components of $\mathcal{W}({\bf a})$ is suggested in the following form:
\begin{eqnarray}
\mathcal{N} \equiv \frac{1}{2} \sum_{\bf a} \left( \abs{\mathcal{W}({\bf a})} - \mathcal{W}({\bf a}) \right),
\label{eq:OQnegativity}
\end{eqnarray}
where it has been shown theoretically and experimentally that a polarization of a single qubit system exhibits $\mathcal{N}>0$ \cite{Ryu13,Ryu19}.

A generalization to $N$-partite OQ function is straightforward, which is also defined by the discrete Fourier transform of a composite expectation for $N$-partite system $C({\bf n}^1, \dots, {\bf n}^N)\equiv C({\bf n}^1)\otimes
\cdots\otimes C({\bf n}^N)$ as
\begin{align}
\mathcal{W}({\bf a}^1, \dots, {\bf a}^N) \equiv \frac{1}{2^{NK}} &\sum_{{\bf n}^1, \dots, {\bf n}^N} (-1)^{- {\bf a}^1 \cdot {\bf n}^1 \cdots - {\bf a}^N \cdot {\bf n}^N} \nonumber \\
&\times C({\bf n}^1, \dots, {\bf n}^N),
\label{eq:quasi_N}
\end{align}
where a tuple ${\bf n}^i =(n_{1}^{i}, n_{2}^{i}, \dots, n_{K}^{i})$ represents possible measurement configurations for $i$-th subsystem having $K$ measurement operators and ${\bf a}^i \cdot {\bf n}^i = \sum_{k} a_{k}^{i} n_{k}^{i}$.

\section{Results and Discussion}
\label{sec:results}
\subsection{Entanglement verification for ideal case}
\label{sec:IDEAL}
Let us now characterize the entanglement of given two-qubit states with the marginal OQ method. We first apply our method to an ideal case of arbitrary two-qubit states obtained with a sequential conduction of an
one-qubit rotation and a CNOT operation that are assumed to be free from noises. The results obtained in this section will be used as a baseline for discussion of the ones obtained in a Si DQD platform including how
noises affect the entanglement of two-qubit states.

As an initial state, we consider the separable two-qubit state of $\ket{00} (=\ket{0} \otimes \ket{0})$, where $\ket{0}$ denotes one of eigenstates associated with the Pauli spin operator $\hat{\sigma}_z$. In general, a time-sequential conduction
of an one-qubit rotation and a two-qubit entangling operation can transform the initial state to an arbitrary two-qubit entangled state. For the rotation gate, here we choose $R_x (\alpha)= \cos(\alpha/2) \openone -i \sin(\alpha/2)
\hat{\sigma}_x$ (rotating the input state by an angle of $\alpha$ around the $x$-axis), where $\openone$ is an identity operator and $\hat{\sigma}_x$ is the Pauli spin operator, and for two-qubit entangling gate we consider
the CNOT operation. Then, the output state can be expressed by

\begin{eqnarray}
\ket{\psi (\alpha)} = \cos \frac{\alpha}{2} \ket{00} - i \sin \frac{\alpha}{2}  \ket{11},
\label{eq:2qubit}
\end{eqnarray}
where $0\leq \alpha \leq \pi$ and, when $\alpha$ is $\pi/2$, it becomes one of the Bell-state that is maximally entangled.

Here we employ two Pauli spin operators $\hat{\sigma}_x$ and $\hat{\sigma}_y$ as local measurements, and then, the OQ function for two-qubit systems can be constructed in the form of Equation (\ref{eq:quasi_N})
with $N=2$. Note that the OQ consists of the selective and sequential measurements in time for local systems as the measurement configuration shown in Figure \ref{fig:schm}. In consequence, when the OQ method
is applied to a composite system involving more than two spatially separated subsystems, the OQ function reveals two different types of nonclassicality; temporal and spatial correlations. The quantum entanglement
is the most well-known indicator of the spatial correlation while the temporal correlation is revealed by the quantum superposition of a single subsystem. Since our aim is to characterize the entanglement, we will
investigate only the spatial correlation. The OQ formula in Equation (\ref{eq:quasi_N}) contains both correlations, so we shall deploy a marginal OQ function in the following form:
\begin{align}
\mathcal{W}_{m} ({\bf c}) = \sum_{{\bf a}^1,{\bf a}^2} & \left( \prod_{(j \neq k)=1}^{2}  \delta_2(c_j - a_{j}^{1} + a_{k}^{2}) \right) \nonumber \\
&\times \mathcal{W}({\bf a}^1,{\bf a}^2),
\label{eq:marginalOQ}
\end{align}
where $a_j, c_j \in \{ 0, 1 \}$, and $\delta_2(x)$ becomes 1 if $x$ is a multiple of 2 and 0 otherwise. Here the $\delta_2$ function can be evaluated only with components related to the spatial correlation, $i.e.$, quantum
entanglement. Then, the marginal OQ function consists of the three terms that correspond to the following measurement configurations (except the case that both subsystems conduct an identity measurement): (i) the
subsystem $1$ selects the local measurement operator $\hat{\sigma}_x$ and the subsystem $2$ selects $\hat{\sigma}_y$, (ii) the opposite of the case (i), and (iii) both subsystems conduct sequential measurements
in the order of $\hat{\sigma}_x$ and $\hat{\sigma}_y$. Finally, we compute the following equation to quantify the nonclassicality of a given two-qubit state: 
\begin{eqnarray}
\mathcal{N}_{m} = \frac{1}{2} \sum_{{\bf c}} \left( \abs{\mathcal{W}_m({\bf c})} - \mathcal{W}_m({\bf c}) \right),
\label{eq:MOQnegativity}
\end{eqnarray}
where the case of $\mathcal{N}_{m}>0$ can be used as the indicator of quantum entanglement \cite{Ryu13}.

\begin{figure}[t]
	\centering
	\includegraphics[width=\columnwidth]{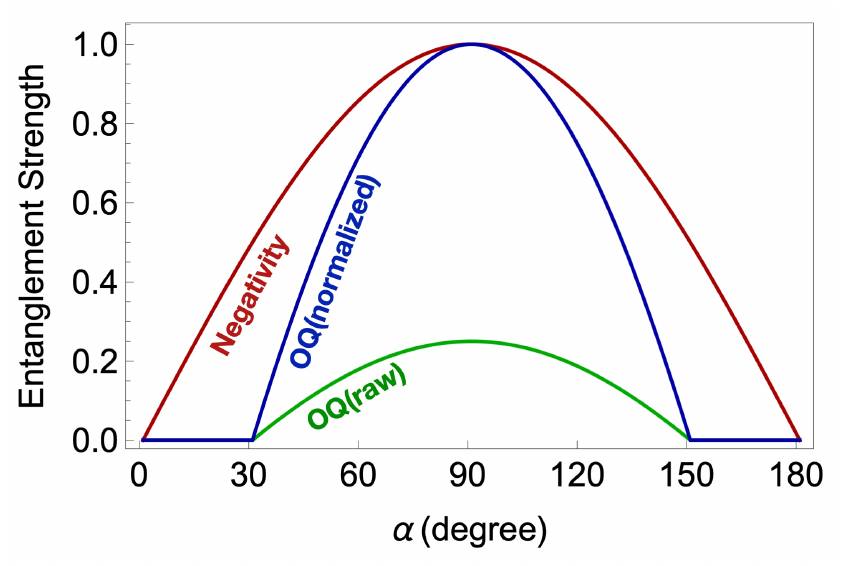}
	\caption{
	\textbf{Comparison of the entanglement detection for the ideal case.} Entanglement strength is plotted as a function of the one-qubit rotation angle $\alpha$. The blue (normalized) and green (raw) lines show the
	results obtained with the marginal OQ method, and the red line indicates the result calculated with the negativity method. The Pauli spin operators $\hat{\sigma}_x$ and $\hat{\sigma}_y$ are used for the marginal
	OQ function. Both methods place the maximal strength at $\alpha=\pi/2$, where the output state becomes a Bell state and therefore is maximally entangled.
	}
\label{fig:OQideal}
\end{figure}

Figure \ref{fig:OQideal} shows the entanglement strength ($\mathcal{N}_{m}$) that is calculated as a function of $\alpha$ for the two-qubit state in Equation (\ref{eq:2qubit}). Here, the green (raw) and the red line indicate
the results calculated with the marginal OQ method and the negativity method that is one of the well-known entanglement measures, respectively. It turns out that the OQ-driven maximum of entanglement strength turns
out to be $0.25$ at $\alpha=\pi/2$ where the output state shown in Equation (\ref{eq:2qubit}) becomes a Bell state that is maximally entangled. In order to compare with the results obtained by the negativity method, we
normalize $\mathcal{N}_{m}$ with respect to its maximal value ($\mathcal{N}_{MAX}$), and, as shown with the blue line in Figure \ref{fig:OQideal}, the normalized result clearly indicates the our method can characterize
the entanglement of the output state quite well when $\pi/6 \leq \alpha \leq 5 \pi/6$. For $\alpha$ that is smaller than $\pi/6$ or greater than $5 \pi /6$, however, the marginal OQ method is not as good as the negativity
method for entanglement detection, and it is because the measurement configurations consisting of the two Pauli spin operators ($\hat{\sigma}_x$ and $\hat{\sigma}_y$) are not sufficient enough to extract the negative
values of the marginal OQ function for the output state. Although this problem can be completely resolved by employing one more spin operator, $\hat{\sigma}_z$ for example, it can be still fair to keep using only two spin
operators in this case, since the result is valid in the range of $\alpha$ where the entanglement becomes significant. In this case, the cost-efficiency that our method has against the negativity method is obvious since the
marginal OQ method involves four measurement configurations while the negativity method requires sixteen. Also, our method is advantageous since the marginal OQ function can be constructed only with directly measurable
quantities that can be easily implemented in reality. Now we move our focus to the realistic DQD system in the section \ref{sec:DQDs}.

\subsection{Entanglement verification for Si DQD system}
\label{sec:DQDs}
\begin{figure*}[t]
	\centering
	\includegraphics[width=\textwidth]{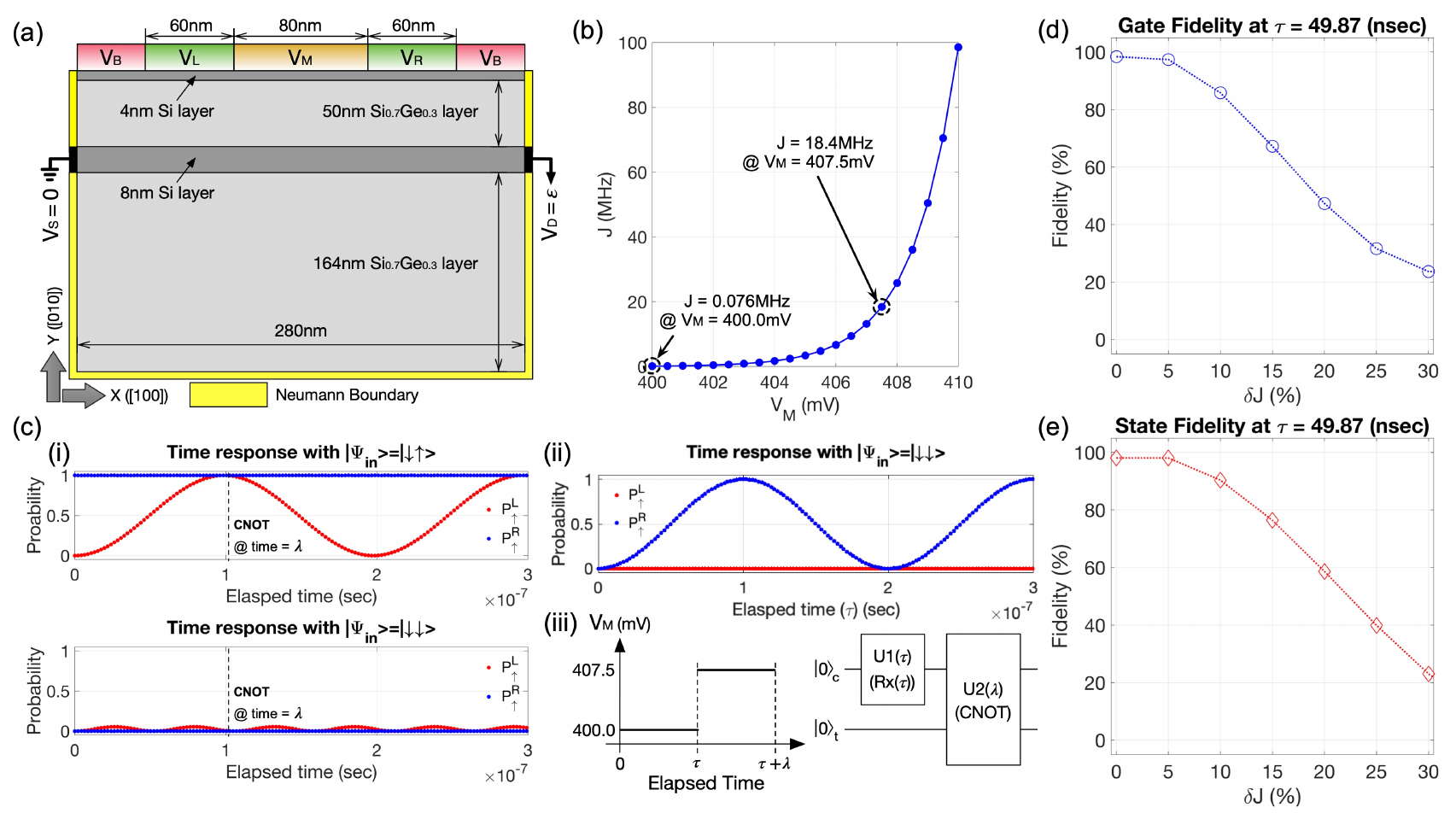}
	\caption{
	\textbf{Simulations of the Si DQD structure.} (a) A 2D simulation domain that is used for our simulations. The real DQD structure \cite{Zajac18} is quite long along the Z([001])-direction, so it is described in a 2D manner with
	a periodic boundary condition along the Z-direction. A bias of 200mV is imposed on the barrier gate ($V_{\text{B}}$), where the source is grounded ($V_{\text{S}}$) and $\delta$ is set to 0.1mV as the drain bias ($V_{\text{D}}$)
	is normally extremely small. A temperature of 1.5K is assumed. (b) $J$ given as a function $V_{\text{M}}$ when $V_{\text{R}}$ = $V_{\text{L}}$ = 555mV. In our case, $J$ $\sim$ 76KHz and 18.4MHz when $V_{\text{M}}$ =
	400mV and 407.5mV, respectively. (c) (i) Two-qubit time response in the regime of strong interaction ($V_M$ = 407.5mV) indicates that the fastest CNOT operation can be achieved in $\sim$1.05$\times$10$^{-7}$ ($\lambda$)
	seconds upon the system intialization. In this case, the gate fidelity of the CNOT operation becomes $98.35\%$. (ii) When the interaction of two QDs is weak ($V_M$ = 400mV), we can make only the right spin oscillate by setting the frequency of AC pulse equal to the Zeeman-splitting energy of the
	right spin. Note that of the label P$^L_\uparrow$(P$^R_\uparrow$) indicate the probability that the up-spin state in the left(right) QD is occupied. (iii) A conceptual illustration for time-dependent control of $V_M$ and resulting
	two-qubit unitary that represents the problem described in the section \ref{sec:IDEAL}. (d, e) The fidelity of the two-qubit unitary and the corresponding output state at $\tau$ = 4.99$\times$10$^{-8}$ seconds (the time spot
	when the output state is maximally entangled) are shown as a function of $\delta J$, which represents the unintentional variation of $J$ with respect to its noise-free value.
	}
\label{fig:QDsim}
\end{figure*}

For entanglement characterization of a more realistic case, we model two-qubit time responses that are generated from a Si DQD structure. Figure \ref{fig:QDsim}(a) illustrates a 2D simulation domain that mimics the reported
DQD structure \cite{Zajac18}, which encodes qubits to electron spins that are created with quantum confinement driven by biases imposed on top electrodes. Bias-dependent electron spins, their Zeeman-splitting and exchange
energy, and corresponding two-qubit time responses are simulated with our in-house simulation tool that employs a multi-scale modeling approach combining semi-classical Thomas-Fermi calculations and electronic structure
simulations coupled to a simple effective mass theory \cite{Wang05}. To initialize the DQD system to a $|$$\downarrow$$\downarrow$$\rangle$ state by filling the ground down-spin state of the left ($|$$\downarrow$$\rangle$$_{\text{L}}$)
and right QD ($|$$\downarrow$$\rangle$$_{\text{R}}$) with a single electron, here we set the left ($V_{\text{L}}$) and right gate bias ($V_{\text{R}}$) to 555mV. The middle gate bias ($V_{\text{M}}$) is set in two ways; 400mV
and 407.5mV where $|$$\downarrow$$\rangle$$_{\text{L}}$ and $|$$\downarrow$$\rangle$$_{\text{R}}$ interact with an exchange energy ($J$) of 76KHz (weak interaction) 18.4MHz (strong interaction) respectively, as shown
in Figure \ref{fig:QDsim}(b). A spatial distribution of the static magnetic field that is generated by a horseshoe-shaped cobalt micro-magnet in the real case \cite{Zajac18}, is taken from the numerical result reported by Neumann
$et$ $al.$ \cite{Neumann15}, and is utilized as an input of simulations. The resulting Zeeman-splitting energy of the left ($E_{\text{ZL}}$) and right spin ($E_{\text{ZR}}$) turn out to be 18.31GHz and 18.45GHz respectively,
and have no remarkable dependency on $V_{\text{M}}$.

Once the system is initialized to a $|$$\downarrow$$\downarrow$$\rangle$ (= $|$$\downarrow$$\rangle$$_{\text{L}}$$\otimes$$|$$\downarrow$$\rangle$$_{\text{R}}$) state, a single-qubit $R_x$($\alpha$) gate and a two-qubit
CNOT gate can be implemented by controlling $V_M$ (and so $J$). When the spin-interaction is weak, two QDs act almost independently and the $R_x$ operation can be achieved by setting the frequency of the AC microwave
pulse to either $E_{\text{ZL}}$ or $E_{\text{ZR}}$ (the phase $\alpha$ is proportional to the duration of the AC pulse ($\tau$)) \cite{Kang21}. In the regime of strong interaction, the single-step CNOT operation can be achieved by
employing the right spin as a control qubit \cite{Russ18}. The two-qubit time responses simulated at $V_M$ = 407.5mV and 400mV are presented in Figure \ref{fig:QDsim}(c)-(i) and Figure \ref{fig:QDsim}(c)-(ii) respectively, revealing that it takes 1.05$\times$10$^{-7}$ seconds ($\lambda$) until the first CNOT operation happens in a noise-free condition. 
In this case, the gate fidelity of the CNOT operation becomes $98.35\%$.
The state in Equation (\ref{eq:2qubit}) then can be obtained in $\tau$+$\lambda$ seconds upon the system
initialization, since we rotate the right spin (at $V_M$ = 400mV) during the first $\tau$ seconds and then conduct a CNOT operation (at $V_M$ = 407.5mV) as illustrated in Figure \ref{fig:QDsim}(c)-(iii). As discussed in the section
\ref{sec:IDEAL}, the output state becomes maximally entangled when $\alpha$ = $\pi/2$ that is achieved at $\tau$ = 4.99$\times$10$^{-8}$ seconds in our simulations. In reality, semiconducting devices cannot be free from charge
noise \cite{Paladino14}, and its simplest model can be incorporated into simulations by introducing unintended fluctuation of exchange interaction \cite{Russ18}. To figure out how charge noise affects the quantum logic operation
shown in Figure \ref{fig:QDsim}(c)-(iii), we change $J$ (the noise-free values: 76KHz at $V_M$ = 400mV and 18.4MHz at $V_M$ = 407.5mV) to $J$$\times$(1+$\delta J$) ($0 \leq \delta J \leq 1$), and calculate the fidelity of the
entire two-qubit circuit and the corresponding output state at $\tau$ = 4.99$\times$10$^{-8}$ seconds. Figure \ref{fig:QDsim}(d) and (e) show the results as a function of $\delta J$, and indicate that both of the gate and state fidelity
start to drop rapidly when the variation becomes larger than 5\% of their noise-free values ($\delta J$ $>$ 0.05).
We note the initial input state here is assumed to be free from noise-driven distortion since the initialization is done when the interaction between QDs is not strong ($V_{\text{M}}$ = 400mV) where the instability of electron
spin states driven by charge noise is not quite remarkable \cite{Ryu22}.

\begin{figure*}
	\centering
	\includegraphics[width=\textwidth]{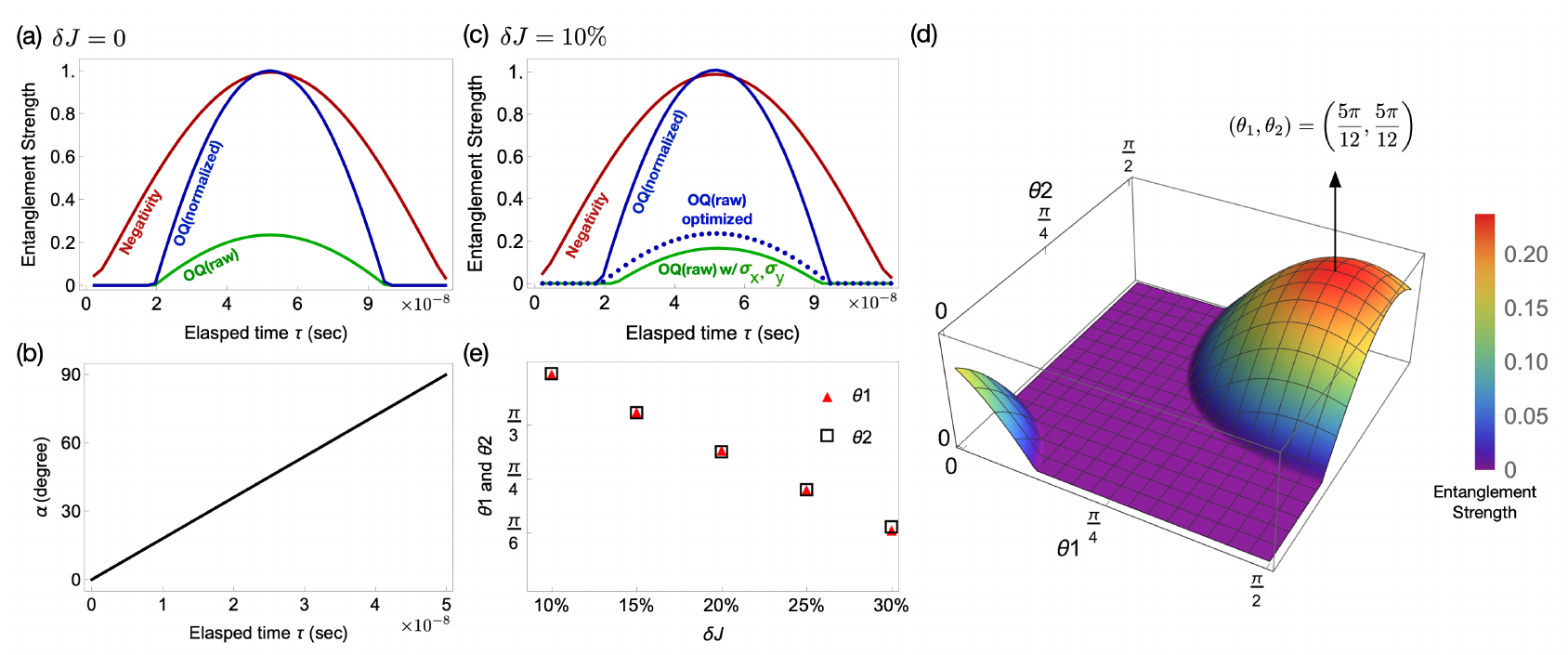}
	\caption{
	\textbf{Entanglement characterization of quantum states generated from the Si DQD structure.} (a) Noise-free entanglement strength ($\delta J=0$) that is calculated with our method (green and blue lines) and the negativity
	method (red line). In both cases, the maximally entangled state is observed at $\tau$ = 4.99$\times$10$^{-8}$ seconds that corresponds a rotation angle ($\alpha$) of $\pi/2$. (b) The correlation between $\tau$ and $\alpha$.
	(c) Entanglement strength at $\delta J=10\%$ that is obtained with the our method with an optimized set of measurement operators (blue point). For the purpose of comparison, the result obtained with $\hat{\sigma}_x$ and
	$\hat{\sigma}_y$ operators are also plotted with an green line. (d) A 3D surface plot showing the optimized set of $\theta_1$ and $\theta_2$ for entanglement characterization when $\delta J$ = 10\%. (e) The optimized values
	of $\theta_1$ and $\theta_2$ are presented as a function of $\delta J$, indicating the precise description of entanglement can be achieved when $\theta_1=\theta_2$.
	}
\label{FIG:DQDj00}
\end{figure*}

Let us now discuss the entanglement characteristic of quantum states that are generated from the Si DQD system. Similarly to the ideal case discussed in the section \ref{sec:IDEAL}, we employ the marginal OQ method with the
two Pauli operators $\hat{\sigma}_x$ and $\hat{\sigma}_y$ to examine the entanglement of time-dependent output states generated by sequential application of a $R_{x} (\alpha)$ and a CNOT operation. The noise-driven characteristic
of quantum entanglement is investigated by changing the noise-free exchange interaction $J$ to $J$$\times$(1+$\delta J$) as we treated to simulate the fidelity shown in Figure \ref{fig:QDsim}(c) and Figure \ref{fig:QDsim}(d).

In Figure \ref{FIG:DQDj00}(a), we show the results of the noise-free case ($\delta J=0$) as a function of the time $\tau$ that is the duration of the AC microwave pulse applied to conduct a $R_{x} (\alpha)$ rotation and is proportional
to the rotation angle $\alpha$ as described in Figure \ref{FIG:DQDj00}(b). The blue (normalized) and green (raw) lines indicate the entanglement strength calculated with the marginal OQ method, and the red line is the one obtained
with the negativity method. Here, the pattern in entanglement strength is generally quite similar to that of the ideal case (Figure \ref{fig:OQideal}), and, in particular, the maximal strength in this case becomes 0.2348 (green line)
at $\tau$ = 4.99$\times$10$^{-8}$ seconds similarly to the ideal case (0.25 at $\alpha$=$\pi/2$). There also exist the intervals of $\tau$ where the entanglement is not precisely characterized with our method. As discussed in the
section \ref{sec:IDEAL}, this is because we employed only two operators ($\hat{\sigma}_x$ and $\hat{\sigma}_y$) to construct the marginal OQ function for cost-efficient calculations. 

Next, we look into what happens in the entanglement characteristic when the platform suffers from noises, $i.e.$, $\delta J \neq 0$. As Figure \ref{fig:QDsim}(d) and \ref{fig:QDsim}(e) show, the noise-driven degradation in fidelity
is quite severe. However, the noise-driven pattern of entanglement characterization does not necessarily follow that of fidelity, and the output state of the noisy two-qubit operation still has meaningful strength of entanglement. Figure
\ref{FIG:DQDj00}(c), which shows the entanglement strength at $\delta J =10\%$, clearly indicates that the result here (blue points) is not quite different from the noise-free one, and both cases have the maximal strength (0.2348
and 0.2331 at $\delta J$ = 0 and 10\%, respectively) at $\tau$ = 4.99$\times$10$^{-8}$. As similarly to the noise-free case, the marginal OQ function here also turns out to be fairly precise so the result normalized to $\mathcal{N}_{MAX}$
(blue line) becomes quite similar to the one obtained with the negativity method (red line). However, it should be noted that the Pauli spin operators employed in the noise-free case do not guarantee the
precise characterization in noisy conditions, so, if the marginal OQ function is constructed with $\hat{\sigma}_x$ and $\hat{\sigma}_y$, the maximal entanglement strength at $\delta J$ = 10\% becomes 0.1669 as the green line
in Figure \ref{FIG:DQDj00}(c) indicates. Due to the degradation in fidelity, the output state at $\tau$ = 4.99$\times$10$^{-8}$ in noisy conditions is not ``numerically'' identical to that of the noise-free case, so the marginal OQ method
coupled to the operators $\hat{\sigma}_x$ and $\hat{\sigma}_y$ does not guarantee precise description of the entanglement strength. In consequence, the general rule in nosy conditions is to find out a set of new operators with
which the two-qubit states can be fairly characterized. 

\begin{figure}[t]
	\centering
	\includegraphics[width=\columnwidth]{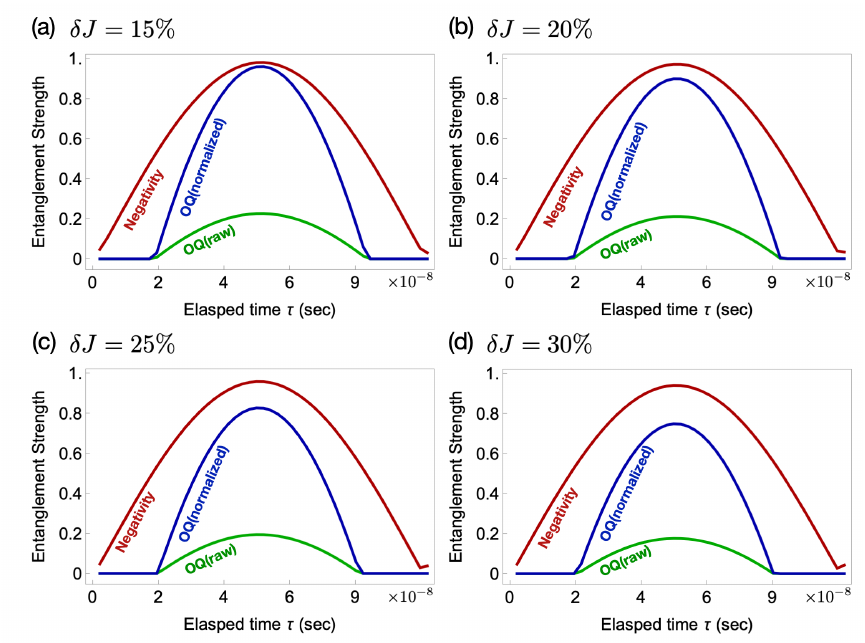}
	\caption{
	\textbf{Variation in the entanglement strength of the two-qubit output state with increasing $\delta J$.} As the Si DQD system suffers from stronger charge noise, the maximal entanglement strength of the output
	state at $\alpha$ = $\pi/2$ keep reducing, so it becomes $0.2252$, $0.2109$ and $0.1757$ when $\delta J$ is increased to 15\%, 20\% and 30\%, respectively (the nose-free case reads 0.2348). The reduction
	in entanglement strength, however, turns out to be much weaker than the case of fidelity that is shown in Figure \ref{fig:QDsim}(d) (gating) and \ref{fig:QDsim}(e) (state).}
	\label{fig:plotjtot}
\end{figure}

To get the operators that are appropriate for description of noisy states, we consider two variables defined by $\hat{\sigma}_1 (\theta_1) = \cos (\theta_1) \hat{\sigma}_{x} - \sin (\theta_1) \hat{\sigma}_{y}$ and $\hat{\sigma}_{2}
(\theta_2) = \sin (\theta_2) \hat{\sigma}_{x} + \cos (\theta_2) \hat{\sigma}_{y}$ ($0 \leq \theta_{1,2} \leq \pi/2$), and numerically determine ($\theta_1$, $\theta_2$) that maximizes the entanglement strength of the output state at
$\tau$ = 4.99$\times$10$^{-8}$ seconds. When $\delta J =10\%$, in particular, both $\theta_1$ and $\theta_2$ are determined as $\sim$5$\pi/$12 as shown in Figure \ref{FIG:DQDj00}(d). We note that, for all the $\delta J$ values
considered in this work, the entanglement is well characterized with the marginal OQ method
when $\theta_1$ = $\theta_2$,
as indicated by Figure
\ref{FIG:DQDj00}(e) where the numerically obtained values of $\theta_1$ and $\theta_2$ are shown as a function of $\delta J$ (corresponding measurement operators are presented in Appendix \ref{app:opt_setting}). By incorporating
these new operators into the marginal OQ method, we calculate the entanglement strength of two-qubit states increasing $\delta J$ up to 30\% with a 5\% step, and show the results in Figure \ref{fig:plotjtot}(a)-(d). Here we find a
general pattern that the maximal entanglement strength at $\tau$ = 4.99$\times$10$^{-8}$ slightly reduces as the DQD platform experiences stronger charge noise (being represented with larger $\delta J$), so its raw magnitude
(green line) reads 0.2252 at $\delta J = 15\%$ but becomes 0.2109 and 0.1757 when $\delta J$ is increased to 20\% and 30\%, respectively.
Our method still retain the accuracy when $\delta J$ $\leq$ 20\%. At
$\delta J$ larger than 20\%, however, the OQ starts to underestimate the entanglement strength so, when $\delta J$ = 30\%, the normalized result (blue line) becomes $\sim$20\% smaller in magnitude than the values
calculated with the negativity method (red line). Note that 
for the negativity method, the maximal strength becomes $0.9805$, $0.9704$, and $0.9398$ when $\delta J$ is increased to 15\%, 20\% and 30\%, respectively. For the same $\delta J$ cases, the OQ normalized values read $0.9591$, $0.8982$, and $0.7483$.

At this point, it should be noted that the noise-driven variation in entanglement strength here delivers an important message. As shown in Figure \ref{fig:QDsim}(d) and \ref{fig:QDsim}(e), the gate and state fidelity are sensitive
to charge noises, and are sharply reduced as $\delta J$ increases. But Figure \ref{fig:plotjtot} clearly shows that the noise-driven ($i.e.,$ $\delta J$-carried) variation in entanglement strength of the two-qubit state is generally much
weaker than what the fidelity shows, so, at $\delta J$ = 30\%, the state fidelity drops to $\sim$20\% while a significant portion ($>$ 70\%) of the entanglement resource is still retained. Additionally, in spite of the underestimation
that is observed when the noise is too strong, the results we present can be still fairly solid enough to claim the utility of the marginal OQ method as a cost-efficient indicator of entanglement strength, where the cost-efficiency
of our method against the negativity method will sharply increase as the size (in qubits) of targeted quantum states increases.

\section{Remarks}
\label{sec:remarks}
We characterize the entanglement of two quantum bits (qubits) states that are generated in a silicon (Si) double quantum dot (DQD) platform. For arbitrary states that are produced through the sequential conduction of a single qubit
rotation and a controlled-NOT operation, we employ the marginal operational quasiprobability (OQ) method to directly quantify their entanglement resource. Here we show that the marginal OQ function, which can be constructed
solely with directly measurable operators, can serve as a fairly solid indicator of quantum entanglement even though a given state is contaminated too much with noises, since it characterizes the entanglement strength with reasonable
accuracy and lower computing cost compared to the well-known negativity method that involves the full state tomography process. We also rigorously investigate how two-qubit states in a Si DQD system are affected by charge
noises that are omnipresent in semiconductor devices. While we see that the noise drives huge degradation in fidelity, its effect on the entanglement resource turns out to be much weaker so more than 70\% of resource can be
retained for maximally entangled two-qubit states even in a strongly noisy condition where the state fidelity drops to around 20\%.

\begin{acknowledgments}
This work has been carried out under the support of the National Research Foundation of Korea (NRF) grant (NRF-2020M3E4A1079792 and NRF-2022M3K2A1083890).
\end{acknowledgments}

\appendix
\section{Rotating measurement operators}
\label{app:opt_setting}
We here present the optimized measurement operators in detail. For $\delta J=10\%$, the two operators read
\begin{eqnarray*}
\begin{pmatrix}
0 & 0.2669 + 0.9637 i \\
0.2669 - 0.9637 i & 0
\end{pmatrix},
\end{eqnarray*}
\begin{eqnarray*}
\begin{pmatrix}
0 & 0.9618 - 0.2737 i \\
0.9618 + 0.2737 i & 0
\end{pmatrix}.
\end{eqnarray*}

For $\delta J=15\%$,
\begin{eqnarray*}
\begin{pmatrix}
0 & 0.4386 + 0.8987 i \\
0.4386 - 0.8987 i & 0
\end{pmatrix},
\end{eqnarray*}
\begin{eqnarray*}
\begin{pmatrix}
0 & 0.8930 - 0.4501 i \\
0.8930 + 0.4501 i & 0
\end{pmatrix}.
\end{eqnarray*}

For $\delta J=20\%$,
\begin{eqnarray*}
\begin{pmatrix}
0 & 0.5983 + 0.8012 i \\
0.5983 - 0.8012 i & 0
\end{pmatrix},
\end{eqnarray*}
\begin{eqnarray*}
\begin{pmatrix}
0 & 0.7926 - 0.6098 i \\
0.7926 + 0.6098 i & 0
\end{pmatrix}.
\end{eqnarray*}

For $\delta J=25\%$,
\begin{eqnarray*}
\begin{pmatrix}
0 & 0.7398 + 0.6729 i \\
0.7398 - 0.6729 i & 0
\end{pmatrix},
\end{eqnarray*}
\begin{eqnarray*}
\begin{pmatrix}
0 & 0.6667 - 0.7453 i \\
0.6667 + 0.7453 i & 0
\end{pmatrix}.
\end{eqnarray*}

For $\delta J=30\%$,
\begin{eqnarray*}
\begin{pmatrix}
0 & 0.8563 + 0.5163 i \\
0.8563 - 0.5163 i & 0
\end{pmatrix},
\end{eqnarray*}
\begin{eqnarray*}
\begin{pmatrix}
0 & 0.5219 - 0.8530 i\\
0.5219 + 0.8530 i & 0
\end{pmatrix}.
\end{eqnarray*}

\bibliographystyle{quantum}

\begin{thebibliography}{99}

\bibitem{Horodecki09}
Ryszard Horodecki, Pawe\l{} Horodecki, Micha\l{} Horodecki, and Karol
  Horodecki.
\newblock ``Quantum entanglement''.
\newblock \href{https://dx.doi.org/10.1103/RevModPhys.81.865}{Rev. Mod. Phys.
  {\bf 81}, 865--942}~(2009).

\bibitem{ReviewBell}
Nicolas Brunner, Daniel Cavalcanti, Stefano Pironio, Valerio Scarani, and
  Stephanie Wehner.
\newblock ``Bell nonlocality''.
\newblock \href{https://dx.doi.org/10.1103/RevModPhys.86.419}{Rev. Mod. Phys.
  {\bf 86}, 419--478}~(2014).

\bibitem{Teleportation}
Charles~H. Bennett, Gilles Brassard, Claude Cr\'epeau, Richard Jozsa, Asher
  Peres, and William~K. Wootters.
\newblock ``Teleporting an unknown quantum state via dual classical and
  einstein-podolsky-rosen channels''.
\newblock \href{https://dx.doi.org/10.1103/PhysRevLett.70.1895}{Phys. Rev.
  Lett. {\bf 70}, 1895--1899}~(1993).

\bibitem{ShorAlgorithm}
P. W. Shor.
\newblock ``Algorithms for quantum computation: discrete logarithms and
  factoring''.
\newblock In Proceedings 35th Annual Symposium on Foundations of Computer
  Science.
\newblock \href{https://dx.doi.org/10.1109/SFCS.1994.365700}{Pages 124--134}.
\newblock ~(1994).

\bibitem{Changhyoup21}
Changhyoup Lee, Benjamin Lawrie, Raphael Pooser, Kwang-Geol Lee, Carsten
  Rockstuhl, and Mark Tame.
\newblock ``Quantum plasmonic sensors''.
\newblock \href{https://dx.doi.org/10.1021/acs.chemrev.0c01028}{Chemical
  Reviews {\bf 121}, 4743--4804}~(2021).

\bibitem{Arute19}
Frank Arute, Kunal Arya, and Ryan Babbush~$\it{et}$ $\it{al}$.
\newblock ``Quantum supremacy using a programmable superconducting processor''.
\newblock \href{https://dx.doi.org/10.1038/s41586-019-1666-5}{Nature {\bf 574},
  505--510}~(2019).

\bibitem{Mooney19}
Gary~J. Mooney, Charles~D. Hill, and Lloyd C.~L. Hollenberg.
\newblock ``Entanglement in a 20-qubit superconducting quantum computer''.
\newblock \href{https://dx.doi.org/10.1038/s41598-019-49805-7}{Scientific
  Reports {\bf 9}, 13465}~(2019).

\bibitem{Pogorelov21}
I.~Pogorelov, T.~Feldker, Ch.~D. Marciniak, L.~Postler, G.~Jacob,
  O.~Krieglsteiner, V.~Podlesnic, M.~Meth, V.~Negnevitsky, M.~Stadler,
  B.~H\"ofer, C.~W\"achter, K.~Lakhmanskiy, R.~Blatt, P.~Schindler, and
  T.~Monz.
\newblock ``Compact ion-trap quantum computing demonstrator''.
\newblock \href{https://dx.doi.org/10.1103/PRXQuantum.2.020343}{PRX Quantum
  {\bf 2}, 020343}~(2021).

\bibitem{Debnath16}
S.~Debnath, N.~M. Linke, C.~Figgatt, K.~A. Landsman, K.~Wright, and C.~Monroe.
\newblock ``Demonstration of a small programmable quantum computer with atomic
  qubits''.
\newblock \href{https://dx.doi.org/10.1038/nature18648}{Nature {\bf 536},
  63--66}~(2016).

\bibitem{Wright19}
K.~Wright, K.~M. Beck, S.~Debnath, J.~M. Amini, Y.~Nam, N.~Grzesiak, J.~S.
  Chen, N.~C. Pisenti, M.~Chmielewski, C.~Collins, K.~M. Hudek, J.~Mizrahi,
  J.~D. Wong-Campos, S.~Allen, J.~Apisdorf, P.~Solomon, M.~Williams, A.~M.
  Ducore, A.~Blinov, S.~M. Kreikemeier, V.~Chaplin, M.~Keesan, C.~Monroe, and
  J.~Kim.
\newblock ``Benchmarking an 11-qubit quantum computer''.
\newblock \href{https://dx.doi.org/10.1038/s41467-019-13534-2}{Nature
  Communications {\bf 10}, 5464}~(2019).

\bibitem{Watson18}
T.~F. Watson, S.~G.~J. Philips, E.~Kawakami, D.~R. Ward, P.~Scarlino,
  M.~Veldhorst, D.~E. Savage, M.~G. Lagally, Mark Friesen, S.~N. Coppersmith,
  M.~A. Eriksson, and L.~M.~K. Vandersypen.
\newblock ``A programmable two-qubit quantum processor in silicon''.
\newblock \href{https://dx.doi.org/10.1038/nature25766}{Nature {\bf 555},
  633--637}~(2018).

\bibitem{Steger12}
M.~Steger, K.~Saeedi, M.~L.~W. Thewalt, J.~J.~L. Morton, H.~Riemann, N.~V.
  Abrosimov, P.~Becker, and H.-J. Pohl.
\newblock ``Quantum information storage for over 180 s using donor spins in a
  ${}^{28}$SI {\textquotedblleft}semiconductor vacuum{\textquotedblright}''.
\newblock \href{https://dx.doi.org/10.1126/science.1217635}{Science {\bf 336},
  1280--1283}~(2012).

\bibitem{Tyryshkin12}
Alexei~M. Tyryshkin, Shinichi Tojo, John J.~L. Morton, Helge Riemann,
  Nikolai~V. Abrosimov, Peter Becker, Hans-Joachim Pohl, Thomas Schenkel,
  Michael L.~W. Thewalt, Kohei~M. Itoh, and S.~A. Lyon.
\newblock ``Electron spin coherence exceeding seconds in high-purity silicon''.
\newblock \href{https://dx.doi.org/10.1038/nmat3182}{Nature Materials {\bf 11},
  143--147}~(2012).

\bibitem{Veldhorst14}
M.~Veldhorst, J.~C.~C. Hwang, C.~H. Yang, A.~W. Leenstra, B.~de~Ronde, J.~P.
  Dehollain, J.~T. Muhonen, F.~E. Hudson, K.~M. Itoh, A.~Morello, and A.~S.
  Dzurak.
\newblock ``An addressable quantum dot qubit with fault-tolerant
  control-fidelity''.
\newblock \href{https://dx.doi.org/10.1038/nnano.2014.216}{Nature
  Nanotechnology {\bf 9}, 981--985}~(2014).

\bibitem{Veldhorst15}
M.~Veldhorst, C.~H. Yang, J.~C.~C. Hwang, W.~Huang, J.~P. Dehollain, J.~T.
  Muhonen, S.~Simmons, A.~Laucht, F.~E. Hudson, K.~M. Itoh, A.~Morello, and
  A.~S. Dzurak.
\newblock ``A two-qubit logic gate in silicon''.
\newblock \href{https://dx.doi.org/10.1038/nature15263}{Nature {\bf 526},
  410--414}~(2015).

\bibitem{Zajac18}
D.~M. Zajac, A.~J. Sigillito, M.~Russ, F.~Borjans, J.~M. Taylor, G.~Burkard,
  and J.~R. Petta.
\newblock ``Resonantly driven cnot gate for electron spins''.
\newblock \href{https://dx.doi.org/10.1126/science.aao5965}{Science {\bf 359},
  439--442}~(2018).

\bibitem{Otfried09}
Otfried G{\"u}hne and G{\'e}za T{\'o}th.
\newblock ``Entanglement detection''.
\newblock
  \href{https://dx.doi.org/https://doi.org/10.1016/j.physrep.2009.02.004}{Physics
  Reports {\bf 474}, 1--75}~(2009).

\bibitem{Wigner32}
E.~Wigner.
\newblock ``On the quantum correction for thermodynamic equilibrium''.
\newblock \href{https://dx.doi.org/10.1103/PhysRev.40.749}{Phys. Rev. {\bf 40},
  749--759}~(1932).

\bibitem{Husimi40}
K.~Husimi.
\newblock ``Some formal properties of the density matrix''.
\newblock \href{https://dx.doi.org/10.11429/ppmsj1919.22.4_264}{Proceedings of
  the Physico-Mathematical Society of Japan. 3rd Series {\bf 22},
  264--314}~(1940).

\bibitem{Glauber63}
Roy~J. Glauber.
\newblock ``Coherent and incoherent states of the radiation field''.
\newblock \href{https://dx.doi.org/10.1103/PhysRev.131.2766}{Phys. Rev. {\bf
  131}, 2766--2788}~(1963).

\bibitem{Sudarshan63}
E.~C.~G. Sudarshan.
\newblock ``Equivalence of semiclassical and quantum mechanical descriptions of
  statistical light beams''.
\newblock \href{https://dx.doi.org/10.1103/PhysRevLett.10.277}{Phys. Rev. Lett.
  {\bf 10}, 277--279}~(1963).

\bibitem{Cahill69}
K.~E. Cahill and R.~J. Glauber.
\newblock ``Density operators and quasiprobability distributions''.
\newblock \href{https://dx.doi.org/10.1103/PhysRev.177.1882}{Phys. Rev. {\bf
  177}, 1882--1902}~(1969).

\bibitem{Ferrie2011}
Christopher Ferrie.
\newblock ``Quasi-probability representations of quantum theory with
  applications to quantum information science''.
\newblock \href{https://dx.doi.org/10.1088/0034-4885/74/11/116001}{Reports on
  Progress in Physics {\bf 74}, 116001}~(2011).

\bibitem{Jiyong2015}
Jiyong Park, Junhua Zhang, Jaehak Lee, Se-Wan Ji, Mark Um, Dingshun Lv, Kihwan
  Kim, and Hyunchul Nha.
\newblock ``Testing nonclassicality and non-gaussianity in phase space''.
\newblock \href{https://dx.doi.org/10.1103/PhysRevLett.114.190402}{Phys. Rev.
  Lett. {\bf 114}, 190402}~(2015).

\bibitem{Sperling2018}
J. Sperling and I.~A. Walmsley.
\newblock ``Quasiprobability representation of quantum coherence''.
\newblock \href{https://dx.doi.org/10.1103/PhysRevA.97.062327}{Phys. Rev. A
  {\bf 97}, 062327}~(2018).

\bibitem{Sperling2020}
J Sperling and W~Vogel.
\newblock ``Quasiprobability distributions for quantum-optical coherence and
  beyond''.
\newblock \href{https://dx.doi.org/10.1088/1402-4896/ab5501}{Physica Scripta
  {\bf 95}, 034007}~(2020).

\bibitem{Bohmann2020}
Martin Bohmann, Elizabeth Agudelo, and Jan Sperling.
\newblock ``Probing nonclassicality with matrices of phase-space
  distributions''.
\newblock \href{https://dx.doi.org/10.22331/q-2020-10-15-343}{{Quantum} {\bf
  4}, 343}~(2020).

\bibitem{Jiyong2021}
Jiyong Park, Jaehak Lee, Kyunghyun Baek, and Hyunchul Nha.
\newblock ``Quantifying non-gaussianity of a quantum state by the negative
  entropy of quadrature distributions''.
\newblock \href{https://dx.doi.org/10.1103/PhysRevA.104.032415}{Phys. Rev. A
  {\bf 104}, 032415}~(2021).

\bibitem{Ryu13}
Junghee Ryu, James Lim, Sunghyuk Hong, and Jinhyoung Lee.
\newblock ``Operational quasiprobabilities for qudits''.
\newblock \href{https://dx.doi.org/10.1103/PhysRevA.88.052123}{Phys. Rev. A
  {\bf 88}, 052123}~(2013).

\bibitem{Jae17}
Jeongwoo Jae, Junghee Ryu, and Jinhyoung Lee.
\newblock ``Operational quasiprobabilities for continuous variables''.
\newblock \href{https://dx.doi.org/10.1103/PhysRevA.96.042121}{Phys. Rev. A
  {\bf 96}, 042121}~(2017).

\bibitem{Ryu19}
Junghee Ryu, Sunghyuk Hong, Joong-Sung Lee, Kang~Hee Seol, Jeongwoo Jae, James
  Lim, Jiwon Lee, Kwang-Geol Lee, and Jinhyoung Lee.
\newblock ``Optical experiment to test negative probability in context of
  quantum-measurement selection''.
\newblock \href{https://dx.doi.org/10.1038/s41598-019-53121-5}{Scientific
  Reports {\bf 9}, 19021}~(2019).

\bibitem{Kang21}
Ji-Hoon Kang, Junghee Ryu, and Hoon Ryu.
\newblock ``Exploring the behaviors of electrode-driven si quantum dot systems:
  from charge control to qubit operations''.
\newblock \href{https://dx.doi.org/10.1039/D0NR05070A}{Nanoscale {\bf 13},
  332--339}~(2021).

\bibitem{Ryu22}
Hoon Ryu and Ji-Hoon Kang.
\newblock ``Devitalizing noise-driven instability of entangling logic in
  silicon devices with bias controls''.
\newblock \href{https://doi.org/10.1038/s41598-022-19404-0}{Scientific Reports {\bf 12},
  15200}~(2022).

\bibitem{Wang05}
Jing Wang, A.~Rahman, A.~Ghosh, G.~Klimeck, and M.~Lundstrom.
\newblock ``On the validity of the parabolic effective-mass approximation for
  the $\it{I}$-$\it{V}$ calculation of silicon nanowire transistors''.
\newblock \href{https://dx.doi.org/10.1109/TED.2005.850945}{IEEE Transactions
  on Electron Devices {\bf 52}, 1589--1595}~(2005).

\bibitem{Neumann15}
R.~Neumann and L.~R. Schreiber.
\newblock ``Simulation of micro-magnet stray-field dynamics for spin qubit
  manipulation''.
\newblock \href{https://dx.doi.org/10.1063/1.4921291}{Journal of Applied
  Physics {\bf 117}, 193903}~(2015).

\bibitem{Russ18}
Maximilian Russ, D.~M. Zajac, A.~J. Sigillito, F.~Borjans, J.~M. Taylor, J.~R.
  Petta, and Guido Burkard.
\newblock ``High-fidelity quantum gates in si/sige double quantum dots''.
\newblock \href{https://dx.doi.org/10.1103/PhysRevB.97.085421}{Phys. Rev. B
  {\bf 97}, 085421}~(2018).

\bibitem{Paladino14}
E.~Paladino, Y.~M. Galperin, G.~Falci, and B.~L. Altshuler.
\newblock ``${1}/{f}$ noise: Implications for solid-state
  quantum information''.
\newblock \href{https://dx.doi.org/10.1103/RevModPhys.86.361}{Rev. Mod. Phys.
  {\bf 86}, 361--418}~(2014).

\end{thebibliography}

\end{document}